\documentclass[12pt]{iopart}

\usepackage{graphicx}
\usepackage{times}
\newcommand{\al}{\alpha}

\newcommand{\f}{\frac}

\begin{document}

\title[]{Entropy \& viscosity bound of strange stars}

\author{Sibasish Laha$^{1,\dagger}$, Taparati Gangopadhyay$^{1,\dagger}$, Manjari Bagchi$^{1,\dagger}$,
Mira Dey$^{1,*,\dagger}$,\ Jishnu Dey$^{1,*,\ddagger}$, Monika
Sinha$^2$ \& Subharthi Ray$^3$}

\address{$^1$ Dept. of Physics, Presidency College, Kolkata 700 073, India.\\
$^2$ Dept. of Physics, Burdwan Women's  College, Burdwan, West
Bengal, India.\\
$^3$ Inter University Centre for Astronomy and
Astrophysics, Post Bag 4, Ganeshkhind, Pune 411007, India.\\
$^*$ Associate, IUCAA, Pune, India.\\
$^\dagger$ Supported by DST Ramanna grant, Govt. of India.\\
$^\ddagger$ CSIR Emeritus Professor.}

\begin{abstract}

At finite temperature (T) there is a link with general
relativity and hydrodynamics that leads to a lower bound for the
ratio of shear viscosity and entropy density ( $\eta/s$ ). We
find that the bound is saturated in the simple model for quark
matter that we use for strange stars at $T~=~80~MeV$, at the
surface of a strange star. At this $T$ we have the possibility of
cosmic separation of phases. We find that, although strongly
correlated, the quark matter at the surface of strange stars
constitute the most perfect interacting fluid permitted by
nature. At the centre of the star, however, the density is higher
and conditions are more like the results found for perturbative
QCD.

\end{abstract}

\maketitle

\section{Introduction}\label{sec:intro}

We find that the ratio of the kinetic viscosity to entropy
density of strange stars saturates the lowest possible bound.
This is as perfect as an interacting fluid can be. Although it is
not directly relevant to us - the background for the viscous
bound conjecture of Kovtun, Son and Starinets \cite{kss} (KSS) will be
briefly touched upon in this section because of its general
interest. We will call this the KSS bound.

It is popularly known that black holes are endowed with
thermodynamics. In higher dimensional gravity theories there
exist solutions called black branes. They are black holes with
translationally invariant horizons. For these solutions
thermodynamics can be extended to hydrodynamics, the theory that
describes long wavelength deviations from thermal equilibrium. In
the holographic principle, where a black brane corresponds to a
certain finite-temperature quantum field theory in fewer number
of space time dimensions, and the hydrodynamic behaviour of black
brane horizon is identical with the hydrodynamic behaviour in a
dual theory.

The relevant arguments of KSS~\cite{kss} for a generalization of
the viscous bound $4~\pi~\eta/s~>~ 1$, is very interesting since
it only invokes general principles like Heisenberg uncertainty
relation for the typical mean free time of a quasi-particle and
the entropy density $s$ which in turn is proportional to the
density of the quasi particles. From here to our model is just one
short step of identifying the quasi-particles to be the dressed
quarks of a mean field description for a large colour effective
theory. We describe the model in section~\ref{sec:tstar} for the
sake of completeness, emphasizing the possible observational
checks of the model. In section~\ref{sec:calc} we describe the
simplest possible calculation of the viscosity known to all and
compare our results with other calculations. We present a brief
discussion in section~\ref{sec:disc} and finally we present a
summary and conclusion in section~\ref{sec:sumcon}.


\section{Strange stars at finite T}\label{sec:tstar}

The chiral symmetry restoration (CSR) of our model is
represented by an ansatz for density dependent quark mass :
\begin{equation}
M_i = m_i + M_Q ~sech\left(\frac{n_B}{N n_0}\right), \;\;~~~ i =
u, d, s. \label{eq:qm}
\end{equation}
where $n_B = (n _u+n _d+n _s)/3$ is the baryon number density,
$n_0 = 0.17~fm^{-3}$ is the normal nuclear matter number density,
and $N$ is a  parameter.  

The number density for the strange star in our model changes from
the surface where it is between four and five times the normal
nuclear matter number density of $n_0~=~0.17~fm^{-3}$ to about 15
times $n_0$ at the centre.
At high $n_B$ the quark mass $M_i$ falls from a large value $M_Q$
to its current one $m_i$ which we take to be \cite{D98,brdd1}:
$m_u = 4 \;MeV,\; m_d = 7 \;MeV,\; m_s = 150 \;MeV$ and $M_Q \sim 345
~MeV$. Possible variations of the CSR can be incorporated in the
model through $N$.  
The parameters of the modified Richardson potential with different
scales for confinement (350 $MeV$) and asymptotic freedom
(100 $MeV$) has been used to fit the octet and decuplet masses
and magnetic moments \cite{man1, man2}.

The calculation involves a density and temperature dependent
gluon screening and thermal single particle Fermi functions with
the modified two body quark - quark interaction. Along with the
constraints of $\beta $ - equilibrium and charge
neutrality in these calculations, it is found  that energy
minimum occurs at a density $\sim$ 4 to 5 times the normal
nuclear density $n_0$ till $T~=~80~MeV$. This is a relativistic,
self consistent,  mean field calculation. Strange quark matter is
thus self bound by strong interaction itself. The energy density
and pressure of this matter lead to strange quark star through
the Tolman Oppenheimer Volkov (TOV) equation with mass and radius depending on the central
density of the star. The results of the calculation as well as
references to the many application of the model to astrophysical
observations are given in Bagchi, et al. \cite{brdd2}.

\section{Calculations \& Results}\label{sec:calc}

We have done the simplest possible calculation of the viscosity
of a fluid using the expression based on a calculation by
Clausius in 1860:
\begin{equation}
\eta=\frac{1}{3} mv n \lambda \label{kin}
\end{equation}
where $m$ is the mass of the particles and $n$ is the number density.
The mean free path $\lambda$ is given by
\begin{equation}
 \lambda = \f{3}{4 \pi d^2 n}\label{freepath}
\end{equation}
where $d$ is the interaction diameter. The interaction radius
$r~(=d/2)$ is calculated by assuming that the relevant particles
occupy an effective volume $\frac{4}{3}\pi~r^3$ :
\begin{equation}
 r = \left[\frac{3}{4 \pi n}\right]^{1/3}\label{rad}
\end{equation}
In quark matter, $n = n _u+n _d+n _s$. We need to specify the
momentum as follows :
\begin{equation}
m v = \sum_{i = u,d,s} m_iv_i
\end{equation}

\begin{equation}
 m_iv_i = \f{\int_0^\infty k^3 f(k, U_i)dk}{\int_0^\infty k^2 f(k, U_i)dk}\label{mom}
\end{equation}
where the Fermi distribution is
\begin{equation}
 f(k,U_i) = \f{1}{1~+~exp[(U_i - \mu_i)/T]}\label{fermi}
\end{equation}


\begin{figure}
\resizebox{\hsize}{!}{\includegraphics{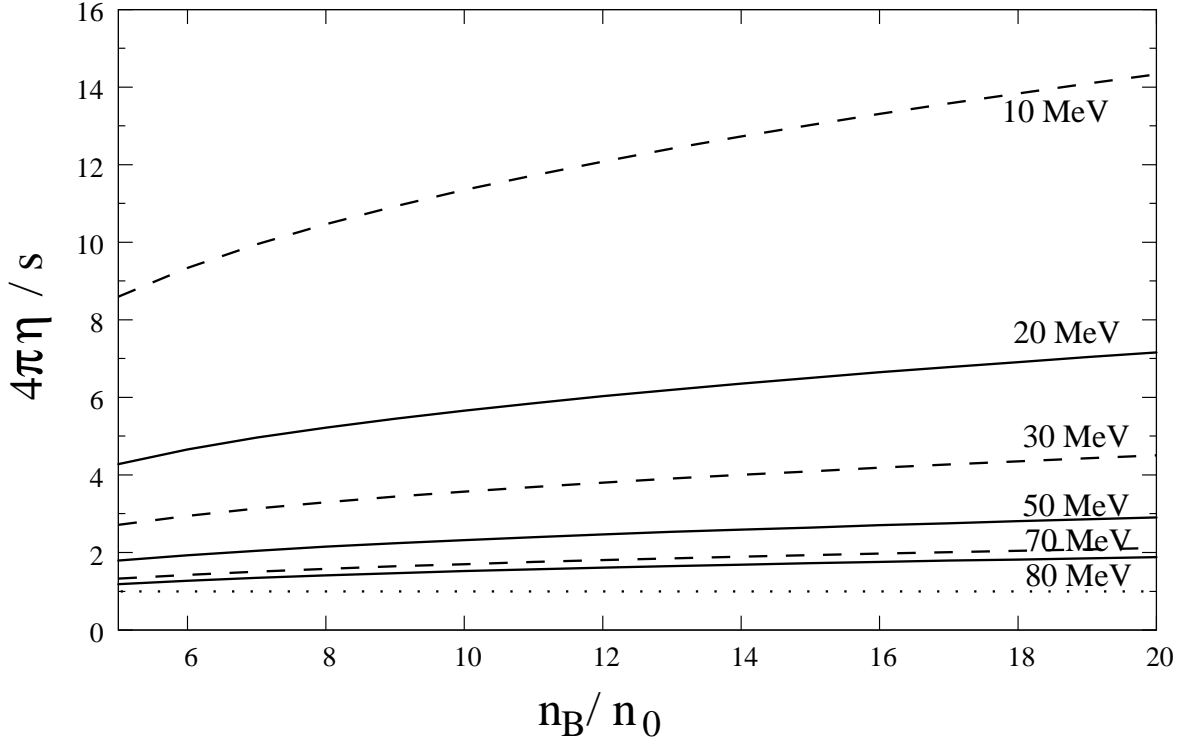}}
\caption{Variation of $\eta /s $ with number density.}
\label{fig:etasvsnum}
\end{figure}

In Fig.(\ref{fig:etasvsnum}) we find that at the surface of the
strange star the KSS bound is saturated. In the model the strange
star is self bound at a certain number density where a
confinement to deconfinement transition takes place due to strong
Debye screening. It is rather satisfying to see that a very
simple evaluation of the viscosity and the entropy in the model
leads to the KSS bound. One can envisage corrections to this
scheme - for example a relativistic relative velocity correction
at the upper limit can give a correction of 30\% in a naive
estimate. But the relative velocity range is from zero and for
lower values the estimate gives smaller numbers.

The variation of $\eta/s$ with the coupling is counter intuitive
as emphasized by Kovtun, Son and Starinets \cite{kss}. We wanted
to check that the ratio in fact increases with decreasing
coupling. To do this we needed the relevant $\al_s$ at each
density.

We have extracted the density dependent strong coupling constant
 $\al_s$ from the density dependent mass ansatz \cite{rdd}. This
is due to the simplified Schwinger-Dyson formalism of Bailin,
Cleymans and Scadron \cite{bcs} using the Dolan-Jackiw Real time
propagator for the quark.

\begin{equation}
\al_s(r,n) = \f{m_{dyn}-M_d(r,n)\pi}{2~
m_{dyn}~ln[\f{\mu(r,n)+\{\mu(r,n)^2-M_d(r,n)^2\}^{1/2}}{M_d(r,n)}]}
\label{dyson}
\end{equation}

We repeat the calculation here for the latest parameter sets
\cite{brdd1} but essentially there is no fundamental change in
$\al_s$, the variation being from 0.5 at low number density at the
star surface to about 0.2 at the highest density in the centre. We
find that $\eta / s $ is a decreasing function of $\alpha_s$ as
discussed for example  by Stephanov \cite{stephanov}. The RHIC is
looking for this region of $\alpha_s$, i.e. large
coupling-nonperturbative region. We see that in our case this
happens at the surface of the strange star.

\begin{table}
\caption{\label{tb:alpha}Variation of the strong coupling constant
$\alpha_s$ with increasing number density predicted by our quark
mass ansatz.}
\begin{indented}
\item[]\begin{tabular}{@{}llllllllll}
\br
$n_B/n_0$ & 4&5&6&7&8&9&10&11&12 \\
\mr $\alpha_s$ &0.522 &0.532  &0.515 &0.486 &0.454 &0.422 &0.393
&0.365 &0.341 \\
\br $n_B/n_0$ & 13&14&15&16&17&18&19&20& \\
\mr $\alpha_s$ &   0.319 &0.299 & 0.281 & 0.265 &0.251 &0.238 &0.226 & 0.215&\\
\br
\end{tabular}
\end{indented}
\end{table}

\begin{figure}
\resizebox{\hsize}{!}{\includegraphics{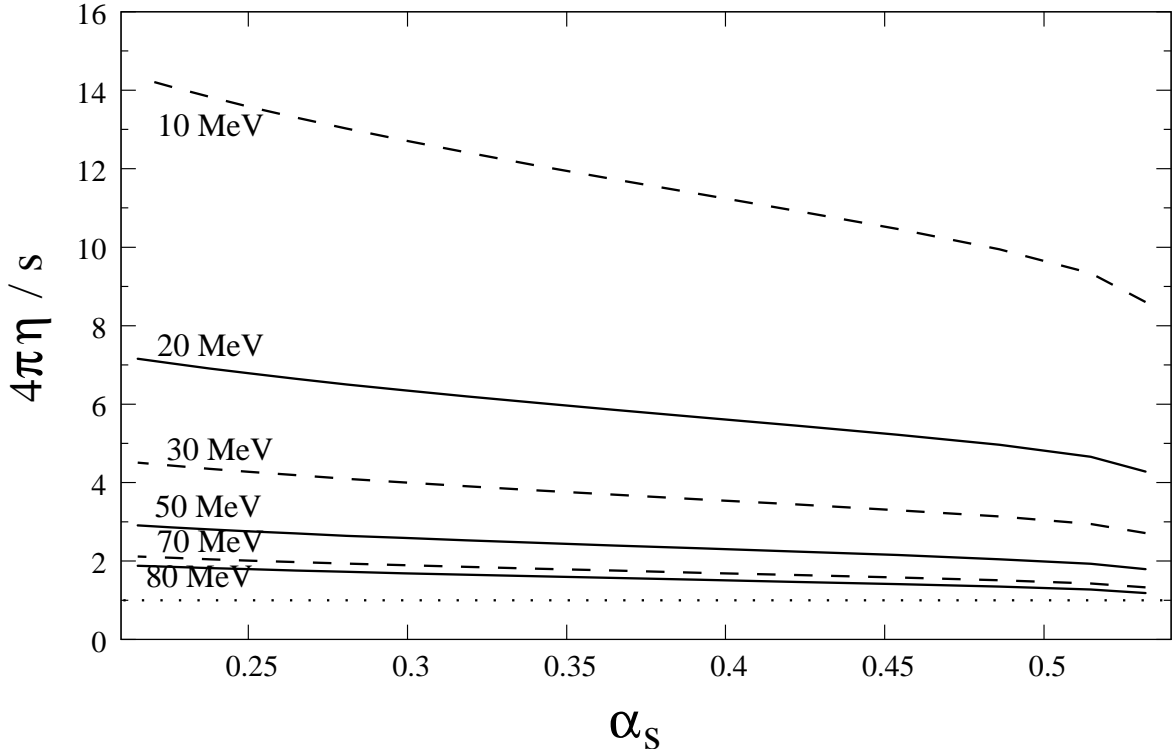}}
\caption{Variation of $\eta /s $ with strong coupling strength.}
\label{fig:etasvsalpha}
\end{figure}

\section{Discussions}\label{sec:disc}

Although Clausius's equation of mean free path is essentially
non-relativistic, we can show that the factor 4/3 in the
expression of relative velocity Clausius's expression will be
changed by a small amount. The factor becomes 1.328, 1.288, 1.218,
1.133 and 1.043 for $\beta~(v/c)$ values of 0.1, 0.3, 0.5, 0.7
and 0.9 respectively. In our model, $\beta$ lies within 0.5-0.7.
So we have neglected this relativistic correction.

We have used the classical expressions for viscosity and mean
free path and treated them semi-classically to reach the
non-perturbative regime of QCD. On the other hand, Heiselberg and
Pethick~\cite{heisel} deduced an expression for viscosity by
evaluating momentum relaxation rate and their applications to
transport processes in degenerate quark matter within
perturbative QCD limit.  This is a variation on the previous
calculation of  Haensel and Jerzak~\cite{HJ}. The expression of
Heiselberg and Pethick is quoted by several other authors~\cite{dorota, madsen}. 
We used the viscosity expression of the
reference~\cite{heisel} and find that at $T~=~80~MeV$ ,  $n_B /
n_0 ~=~ 5$ and  $~ 15$, the $4 \pi \eta /s~=~3.85$ and $ 166.45 $
respectively. It increases with increasing density and decreasing
temperature. For example, at $T~=~50~MeV$, $4 \pi \eta
/s~=~13.99$  and $~597.54$ for $n_B / n_0 ~=~ 5$ and   $~ 15$
respectively.

In a recent paper Lacey has given a very lucid and colourful
representation of viscosity bound for different fluids which we
summarize here. As the ${(T-T_c)}/{T}$ varies from 0.5 to 0,
$\eta/s$ in (a) meson gas goes from 1.2 to 0.4, (b) water goes
from 3.8 to 2.2 (c) liquid nitrogen from 3.4 to 0.8 and (d)
liquid helium from 3.4 to 0.8. The matter in the strange star
seems to be the first so called perfect interacting liquid where
bound reaches the fraction ${(4\pi)}^{-1}$ and thus it may be the
same fluid which Lacey marks as RHIC which stands for
relativistic heavy ion collisions~\cite{lacey}.

\section{Summary and Conclusions}\label{sec:sumcon}

We have found that $\eta / s$ increases with increasing number
density and with decreasing $\alpha_s$. The transport here is
radial hence the kinetic viscosity is the shear viscosity. At the
surface of all strange stars we have an energy density which
makes the pressure zero. This zero pressure is what makes  the
star self bound. And it is the strong interaction which is
responsible zero pressure. The quark matter at the surface looks
like a strongly correlated self bound system where $4 \pi \eta /
s~\sim~1$. It cannot be a mere co-incidence that such our simple
model leads to such an interesting result, connecting the zero
pressure energy density with that of RHIC. Further debate may be
possible as to why it happens, but it is beyond the scope of this
paper.

In summary we find that the strange star surface where the
pressure is zero also turns out to be the region where the KSS
viscosity bound is saturated, and perhaps is the same matter
observed in RHIC.


\ack{
The authors TG, MB, MD and JD are grateful to IUCAA, Pune, and
HRI, Allahabad, India, for short visits. We are grateful to
Rajesh Gopakumar for drawing our attention to the paper by Kovtun,
Son and Starinets and for illuminating discussions.
}


\section*{References}
\begin{thebibliography}{}
\bibitem{kss} Kovtun, P. K., Son, D. T. and Starinets, A. O., 2005, Phys.
Rev. Lett. 94, 111601.

\bibitem{D98} Dey M., Bombaci I.,  Dey J., Ray S. \& Samanta B.C.
1998, Phys.  Lett. B 438, 123.


\bibitem{brdd1} Bagchi, M., Ray, S., Dey, M., Dey, J., 2006, Astron. \& Astrophys. 450, 431.


\bibitem{man1} Bagchi, M., Daw, S., Dey, M., Dey, J., 2004, Nucl. Phys. A 740, 109.

\bibitem{man2} Bagchi, M., Daw, S., Dey, M., Dey, J., 2006, Europhys.
Lett.75, 548.

\bibitem{brdd2} Bagchi, M., Ray, S., Dey, M., Dey, J., 2006, Mon. Not. R. Astron. Soc. 368, 971.

\bibitem{rdd} Ray, S., Dey J., Dey M., 2000, Mod. Phys. Lett. A 15, 1301.

\bibitem{bcs} Bailin, D., Cleymans, J., Scadron, M. D., 1985, Phys. Rev.
D 31 164.

\bibitem{stephanov} Stephanov, M., hep-lat/0701002.

\bibitem{heisel} Heiselberg, H., Pethick, C. J., 1993, Phys. Rev. D
48, 2916.

\bibitem{HJ} Haensel P., Jerzak A. J., Acta. Phys. Pol. 1989, B 20, 141.

\bibitem{dorota} Gondek-Rosinska, D., Gourgoulhon, E., Haensel P., Astron. Astrophys., 2003, 412,
777.

\bibitem{madsen} Madsen, J., Phys. Rev. Lett, 1998, 81, 3311.

\bibitem{lacey} Lacey, R. A., nucl-ex/0701026

\end{thebibliography}
\end{document}